\documentclass{entcs} 
\usepackage{entcsmacro}

\usepackage{latexsym,amssymb,amsmath}
\usepackage{epic, eepic,wrapfig}
\usepackage{subfigure}
\usepackage{luca}           
 
\def\zing{{\sc Zing}}
\def\ctl{{\sc Ctl}}

\newcommand{\Comment}[1]{}

\newcommand{\tcc}{{\sc Tic}}

\begin{document}
\def\lastname{Faella, Legay}
\begin{frontmatter}
  \title{Some Models and Tools for Open Systems} 
  \author{Marco Faella\thanksref{luca}}
\address{Dipartimento di Scienze Fisiche\\
         Universit\`a di Napoli ``Federico II'', Italy}
  \address{
School of Engineering\\ 
University of California, Santa Cruz, USA} 
  \author{Axel Legay\thanksref{fria}}
  \address{Department of Computer Science\\
    University of Li\`ege, Belgium}
  \address{
School of Engineering\\ 
University of California, Santa Cruz, USA}
\thanks[luca]{Supported by the NSF CAREER award
CCR-0132780, and by the ARP award TO.030.MM.D.}
\thanks[fria]{Supported by a F.R.I.A grant.}
\begin{abstract} 
In computer science, there is a distinction between closed systems,
whose behavior is totally determined in advance, and open systems,
that are systems maintaining a constant interaction with an
unspecified environment. 
Closed systems are naturally modeled by transitions systems.
Open systems have been modeled in various ways, including
process algebras, I/O automata, ``modules'', and interfaces.
Games provide a uniform setting in which all these
models can be cast and compared.

In this extended abstract,
we discuss the features and costs related to the game-based
approach to open systems, referring to some of the existing
models.
Finally, we describe a new model of interface, 
called \emph{sociable interface}, 
which is geared towards easier specification,
improved reusability of models, and efficient symbolic implementation. 
\end{abstract}
\end{frontmatter}

\sloppy



\section{Introduction}


In computer system design, there is a distinction between {\em closed\/}
and {\em open\/} systems. A closed system is a system whose behavior is
totally determined by the state of the system, while an open system
interacts with its environment and its behavior
depends on those interactions.
In practice, open systems are those that explicitly distinguish
between internal choice, often represented by input events,
and external choice, represented by output events.

Closed models are often used for specifying and verifying properties
of systems whose behavior is totally known in advance. 
Open models are used to analyze systems that
maintain an ongoing interaction with their environment, such as
embedded systems and control systems. 
Moreover, open systems are suitable to represent components
of larger systems, thus supporting compositional verification.
In that case, inputs are supposed to be provided
by the other components.

Closed systems are naturally modeled as transition systems.
Basically, a
transition system consists of a set of system states, and a set of
transitions between the states. Transitions systems are naturally
nondeterministic, which allows them to represent uncertainty or abstraction
in the system model.

Several successful approaches for the design and verification of
open systems are based on extensions of transition systems
(for example, \cite{Lynch-Tuttle88}). 
In a series of recent works~\cite{luca-taormina03,IA-Marktoberdorf04}, 
de Alfaro et al. argue that
\emph{games} constitute a more natural model for open systems, 
not only for design and verification, but also for refinement and composition.
Notice that in compositional verification,
refinement and composition play a much more central role
than in the traditional verification of closed systems.

Games generalize transition systems by providing a model for multiple
independent sources of nondeterminism. Each source is represented by a
player whose moves correspond to the choices available to that source. 
In particular, two-players games constitute
an expressive model for open systems which allows the distinction
between choices that originate within the system
(output moves of the Output player), and the choices of the
environment (input moves of the Input player).

Game models have been widely used to analyze and solve control
problems for open system. The game view has also been used in the
specification and verification of the interaction between components
\cite{ALW}, and the interaction between components and their environment
\cite{KVW01}. 

In this extended abstract, we discuss the features and the costs
associated with fully game-based models for open systems 
(Section \ref{why-a-game-based-model-?}). 
We then make a short
survey of recent works on open
systems, focusing on communication paradigms,
refinement and tool support (Section \ref{sec-literature}).
We consider those frameworks that are more related to
the interface models of de Alfaro et al.,
which are themselves the object of Section~\ref{interface-models}.
Within Section~\ref{interface-models}, we also describe a new
interface model, called \emph{sociable interfaces},
which is geared towards easier specification,
improved reusability of models, and efficient symbolic implementation.

\section{Game-based Models}
\label{why-a-game-based-model-?}

In this section, we briefly survey the features of game-based 
models for design, verification, refinement, and composition
of open systems. A more complete discussion
can be found in~\cite{luca-taormina03}.

\subsection{Design}
In contrast to many transition system-based approaches such as
\cite{Lynch-Tuttle88}, a game model needs not be \emph{input-enabled}. 
This means that at every state, some inputs can be
illegal. By not accepting certain inputs, the model allows to express
the assumption that the environment never generates these inputs (such
an approach is often referred to as being ``optimistic''). In
this way, environment assumptions can be used to encode restrictions
on the order of the method calls, and the types of return values and
exceptions.
Moreover, the ability to forbid inputs removes the need
to specify ``what happens'' when taking an undesirable
input. 
Such a specification has been pointed to as one of the main
drawbacks of input-enabled approaches.


\subsection{Refinement}

The notion of refinement aims at capturing the relation between an
abstract model of a component and a more detailed one, or between a
model expressing a specification and a model describing an
implementation.
Refinement should satisfy the following \emph{substitutability}
condition. If $P$ refines $Q$, then it should be possible to 
replace $Q$ with $P$ in every context and obtain an equivalent system.

When considering transition systems, refinement is often defined as
trace inclusion or simulation. 
Essentially, all the behaviors of the
implementation must also be behaviors of
the specification.
However, such behavior containment is not an appropriate
notion of refinement for open systems,
because it allows the implementation to
accept less inputs than the specification,
thus violating substitutability.

One solution is to impose the input-enabledness of all components.
However, by definition
this restrictions hampers the ability of the model
to express input assumptions.
As a consequence, sometimes (see~\cite{luca-taormina03}),
proving refinement requires to explicitly model the
environment, thus closing the system.

Alternatively, de Alfaro et al. suggest to replace trace
inclusion or simulation with a notion that is contravariant with
respect to inputs and outputs. More precisely, if $P$ refines $Q$, then

\begin{enumerate}
\item
$P$ accepts at least as many input behaviors as $Q$ does. This means that
$P$ can be used in each environment in where $Q$ can be used;
\item
moreover, when $P$ and $Q$ are subject to the same input behavior, $P$
should produce a subset of the output of $Q$.
\end{enumerate}

If $P$ and $Q$ are both modeled as two-player games between an Input player
and an Output player, then the above definition 
leads to using alternating simulation 
or alternating trace inclusion~\cite{CONCUR98AHKV} 
as the refinement relation. 

\subsection{Composition}

Composition is a basic operation for open systems.
Indeed, one of the motivations for open systems is the
ability to combine them in larger units.
Clearly, we expect components to interact,
or communicate, when composed.

A fundamental communication mode is action synchronization.
When two models are composed, if a model has an enabled transition
that emits the action $a$
and the other has an enabled transition that receives the action $a$, 
they will both take their respective transition,
thus synchronizing on $a$.
This mechanism is common to both process algebras (see for instance
Section~\ref{stuck-free}) 
and automata models 
(see Sections~\ref{io-automata} and~\ref{interface-models}).

What changes is the way in which \emph{missed synchronization}
is handled.
%
Consider two components being composed.
One case of missed synchronization occurs
when a component can produce fewer outputs 
than the other one can accept as input.
All models agree that this case does not represent an error.

On the other hand,
assume that a component generates more
outputs than the other one can accept as input.
In some approaches, such as process algebra,
if a component proposes two outputs but the environment
can only accept one of the two, the other output
does not happen, and no error is flagged.
In some application domains, this is not the intended
behavior, and in Section~\ref{stuck-free}, 
a technique is proposed that flags an error when the above situation
occurs.

The I/O automata model of Section~\ref{io-automata}
solves the problem by enforcing the input-enabled principle.
In that case, no problematic missed synchronizations can happen.
However, as it has been pointed out in the
previous sections, such a constraint has a deep influence on the
design and refinement of the system.

Game models, such as the interfaces of Section~\ref{interface-models},
stipulate that output choice takes
precedence over input choice and introduce a notion of
compatibility. 
Consider two components $P$ and $Q$, in one state of the
composition, if $P$ wants to do an output that cannot be
accepted by an input of $Q$, then incompatibility
occurs. While many approaches would consider the two components to be
incompatible in such a case, the interface approach is optimistic,
by expecting the environment to steer away from error states. 
More
concretely, two components are compatible if there exists a way (an
environment) to use the components together, and ensure that the
environment assumptions of both are met.

\subsection{Costs}

It should come as no surprise that the increased expressivity
of game-based models over transition systems comes at a cost.
As far as refinement is concerned,
while trace inclusion is PSPACE-complete, its alternating version is
EXPTIME-complete.
However, both simulation and its alternating version can
be computed in polynomial time~\cite{CONCUR98AHKV}.

As for model checking, game models support more elaborate 
and therefore more costly properties.
For instance, a common type of property for transition systems
is \emph{safety}: checking that the system remains in a safe
area of its state space.
Since game models recognize two types
of non-determinism, 
two types of safety properties can be of interest.
First, safety under all environments, or equivalently
under the worst environment (pessimistic).
Checking such properties is tantamount to checking
safety in transition systems, as both the system and the environment
cooperate in trying to leave the safe area.

Second, safety under the most favorable environment (optimistic).
In this version, the two sources
of non-determinism play in opposite roles:
the system trying to break free of the safe area, 
and the environment
trying to keep the system in it.
Checking this type of properties is in general more complicated than
checking safety in transition systems.

\section{Some Models for Open Systems}
\label{sec-literature}

In this section, we shortly survey some of the existing models for the
design and the verification of open systems. This section is clearly
not exhaustive, and is mainly intended to show the variety of models
that have been proposed to deal with open systems.

\subsection{I/O Automata}
\label{io-automata}



This section is devoted to a short presentation of the well-known {\em
Input/Output automata} model (I/O automata for short)~\cite{Lynch-Tuttle88}.
An I/O automaton is essentially 
a non-deterministic finite-state automaton with actions labeling each
transition. Actions are classified as either {\em
input}, {\em output}, or {\em internal}. 
As usual for open systems, output and
internal actions are issued by the automaton, while input actions are
issued by the environment. 
I/O automata are input-enabled: in each state all inputs are allowed.

I/O automata can be composed to yield other I/O automata. Two I/O
automata communicate by synchronizing on shared actions (i.e. actions
with the same name). If an I/O automaton generates an output action,
then this action is transmitted to all other automata having the same
action as input. 
This composition is similar to the one provided in
CSP in the sense that I/O automata use simultaneous performance of
actions to synchronize components. 
However, it is slightly different
since the synchronization is leaded only by one automaton: the one
that is issuing the output action. Moreover, if two automata
synchronize on the same action, the resulting action is not hidden,
to allow broadcasting.

One cannot compose two I/O
automata if they have a common output action. 
As explained in~\cite{Lynch-Tuttle88}, 
this is to ensure that only one component leads the
composition. However, this restriction, in addition to the fact that
input and output actions of each components are disjoint, restricts the
possible communication patterns between them. 


Since the I/O automata are input-enabled, refinement
can be captured by trace containment or simulation. 
However, sometimes (see~\cite{luca-taormina03}),
proving refinement requires to explicitly model the
environment, thus closing the system.

There are various languages for specifying I/O automata. One of the
main difficulties in the specification is expressing the
input-enabled principle. 
The process-algebraic languages
proposed in \cite{Vaandraager91,Nicola-Segala95} ensure
input-enabledness by filling in {\em default transitions} for missing
input transitions. Those default transitions are either {\em
self-loops} \cite{Vaandraager91}, or lead to a special state (often
called error state) in where the behavior of the environment is
supposed to be random \cite{Nicola-Segala95}. The drawback of such
languages is that sometimes, specifications need transitions that are
self-loops, and other transitions that go to an error state. In
\cite{Stark-Cleavland-Smolka03}, Stark et al. solve the problem by
employing a notion of well-typedness of terms which guarantees that
all well-typed terms are input-enabled. 

Another specification language is the IOA
language, introduced by Lynch in~\cite{Garland-Lynch98}. 
The language allows the designer to express designs at
different levels of abstraction, leading to a low-level description
that can be translated to real code. 
The language is implemented in the {\em IOA toolset} \cite{IOA}.

The I/O automata model is now highly popular for specifying and
verifying distributed algorithms both manually, and with machine
assistance. 
As an example, Nipkow and Slind
implement the I/O automata model in the theorem prover {\em
Isabelle}
to verify communication protocols~\cite{Nipkow-Slind94}. 

It should be mentioned that there exist many variants of I/O
automata such as {\em probabilistic I/O automata}
\cite{Wu-Smolka-Stark94} that are used to describe systems that
exhibit concurrent and probabilistic behaviors, or {\em hybrid I/O
automata} \cite{lynch-Segala-Vaandraager03} that are used to describe
systems with a mixed discrete-continuous semantics.

\subsection{The Ptolemy II Project}
\label{ptolemy}

In the Ptolemy II project \cite{PTOLEMY}, Lee and others focus on
component-based heterogeneous modeling and design, more in particular
on embedded systems. The work done on this project is too vast to be
presented in this section\footnote{The interested reader is redirected
to \cite{PTOLEMY} for many references on the project.}, and we will
only focus on the part that use a game-based approach.

In Ptolemy II, components are called {\em actors} and they
communicate by means of communication channels called {\em
receivers}. The communication is based on a producer/consumer model.
In fact, receivers just provide an interface that contains methods
like \texttt{put} or \texttt{get} to send or receive data from the
channels they represent.

The choice of the communication domain is let to the classes that
implement the receiver interfaces. Among the available communication
domains, one finds CSP (Communicating sequential processes), PN
(Process Networks), SDF (synchronous data flow), DE (discrete event),
etc.  Most of those domains are detailed in~\cite{Lee-Xiong03,PTOLEMY}

Different domains impose different requirements for actors. As an
example, in the SDF domain actors assume that data is always available
when they call \texttt{get}, while in CSP, they wait the data in a
rendez-vous communication style. One of the main challenges here is to
ensure that an actor can work correctly with a receiver.

One of the solutions proposed in Ptolemy II is based on
an action-based interface formalism (see Section~\ref{interface-models})
that is used to describe the
method calls and receive of actors and receivers. More precisely, the
actors and the receivers are modeled as action-based interfaces whose
actions are both the method calls and returns. Then, an actor and a
receiver can work together if their composition is not empty.

Ptolemy II also employs alternating simulation. However, this is not to
check if an implementation refines its specification but to check if a
domain is a {\em sub-type} of an another one. Using alternating
simulation, one can define a partial order relation between domains. 
As it is explained in \cite{Lee-Xiong03}, this partial
order can be used to design actors that can work in multiple
domains.

\subsection{Stuck-free Conformance}
\label{stuck-free}
In two recent works~\cite{Rajamani02,Fournet04},
Rajamani et al. study a form of compatibility theory
for CCS processes called ``stuck-free conformance''.
In this theory, a model is a CCS process, and composition
is the standard parallel composition in this process algebra~\cite{MilnerCCS}.

CCS processes communicate via synchronization on
shared actions.
Multiple processes can share the same actions,
both as input and as output.
However, at any given computation step, 
an output can synchronize
with exactly one corresponding input,
giving rise to an internal action.
The basic communication mode is thus many-to-one.

Given a set of actions $A$, a module is said to be \emph{stuck}
on $A$ if, once communication on $A$ is forbidden
(\emph{restricted} in CCS terminology), 
the module still has pending actions in $A$.
In other words, the module should never try to send
a message to the environment using an action in $A$,
nor it should receive input actions in $A$.
Thus, 
stuck-freedom is only checked once the system has
been closed by the restriction operator.
The theory is enriched by a refinement relation,
called \emph{conformance}, that preserves stuck-freedom.

The conformance check is one of the refinement
relations supported by the tool 
\zing\footnote{\zing\ needs the commercial package ``Visual Studio .NET 2003'' to function.}
\cite{zing04}.
This tool accepts a rich input language
and builds an enumerative finite-state model of it.
Additionally, \zing\ is capable of extracting
its models from common programming languages.
Thus, it is possible to use \zing\ to automatically
check whether a concrete software implementation 
refines its
specification (called \emph{contract} in the \zing\ literature).
In~\cite{Fournet04}, the authors report that the tool
was successfully used in this fashion, leading to the
discovery of several bugs in a distributed software system.

\subsection{Game Semantics}
\label{game-semantics}

In a series of recent papers~\cite{Ghica03,Ghica04},
Abramsky, Ghica and others outline the application
of game semantics~\cite{Abramsky97} to the automatic
analysis and verification of software.
Concurrency is not treated in these works.

The basic idea of game semantics is to interpret
each term in a programming language as a game between
the program and its environment.
For instance the r-value ``$x$'', 
where $x$ is an integer variable,
is interpreted as a game where the program asks
the value of ``$x$'' to the environment,
and the environment answers with an integer.
If such term is inserted in a context where ``$x$'' has been
assigned a certain value ``$c$'', then
the environment will be forced to reply ``$c$''
to the program's question.
So, this formalism allows for a completely compositional semantics.
Moreover, the semantics is fully abstract, in that
programs that are observationally equivalent
correspond to the same game.

A tool has been built to automatically build the
game associated with a program, in the form
of a finite graph, whose edges are
labeled by moves of the program and the environment.
This graph can be interpreted as a labeled transition system
and fed into a traditional model-checker,
to check pessimistic properties of the type 
``for all environments, $\varphi$''.

\subsection{Modular Verification of Features}
\label{features}

Feature-oriented design is a field
where large systems are partitioned in a set
of modules, each of which representing a feature
of the system. 
Features are represented by finite-state machines,
and they are \emph{sequentially} composed 
by linking the final states
of one feature to the initial state of another.

In~\cite{Fisler02,Fisler04}, the authors explore the idea of
treating each feature as an open model,
enabling compositional verification of
\ctl\ properties.

Given a \ctl\ property $\varphi$, algorithms are provided
to generate constraints that the environment 
(the other features)
must satisfy in order for the composed system
to satisfy $\varphi$.
Dually, each feature provides a set of guarantees to
the environment, in the form of 
the so-called \emph{data environment}.

Albeit different in scope, and based on sequential rather
than parallel composition, this theory exhibits the main ingredients
of the interface approach to open systems, which we treat in
the following section.

\section{Interface Models}
\label{interface-models}

Recently, we have proposed a game-based model called 
\emph{sociable interfaces} \cite{luca-frocos05}. 
The state of a sociable interface
consists of a value assignment to a
set of variables.
Variables are partioned into \emph{local} variables,
that are owned by a specific interface,
and \emph{global} ones, that can be updated by any interface.

Synchronization and communication are based on
two main ideas. The first idea is that the same action can appear as a
label of both input and output transitions: when the action labels
output transitions, it means that the interface can emit the action;
when the action labels an input transition, it means that the action
can be accepted if sent from other components.  Depending on whether
the action labels only input transitions, only output transitions, or
both kind of transitions, we have different synchronization schemes.

For instance, if an action $a$ is associated only with output
transitions, it means that the interface can emit $a$, but cannot
receive it, and thus it cannot be composed with any other interface
that emits $a$.  Conversely, if $a$ is associated only with input
transitions, it means that the interface accepts $a$ from other
interfaces, but will not emit $a$.  Finally, if $a$ is associated both
with input and output transitions, it means that the interface can
both emit $a$, and accept $a$ when emitted by other interfaces.

The second idea is that global variables do not belong to specific
interfaces: the same global variable can be updated by multiple
interfaces. In an interface, the output transitions associated with an
action specifies how global variables can be updated when the
interface emits the action; the input transition associated with an
action specifies constraints on how other interfaces can update the
global variables when emitting the action. By limiting the sets of
variables whose value must be tracked by the interfaces, and by
introducing appropriate non-interference conditions among interfaces,
we can ensure that interfaces can participate in complex communication
schemes with limited knowledge about the other participants.  In
particular, interfaces do not need to know in advance the number or
identities of the other interfaces that take part in communication
schemes. This facilitates component reuse, as the same interface model
can be used in different contexts.

We have implemented the theory of sociable interfaces in a tool called
\tcc\ (for \emph{Tool for Interface Compatibility}). For the tool to
handle interesting interfaces, we represent the state-space of an
interface and its transition relations symbolically, i.e. using
MDDs/BDDs.  
The tool takes as input interfaces specified in a guarded-command
style language.
Then, the user has the
options of performing the following operations: \emph{(i)} compose two
interfaces, \emph{(ii)} verify refinement between two interfaces, and
\emph{(iii)} check safety properties of an interface.  All of the
above operations can be computed efficiently using our symbolic
representation.
We are currently working on extending both the theory and
the implementation to discrete real-time systems.

Previous game-based models, such as interface automata
\cite{FSE01-IA,IA-Marktoberdorf04} and interface modules
\cite{EMSOFT01,CAV02-IM} were based on either actions, or variables,
but not both. The rest of this section is devoted to a quick
presentation of such interface models.

\subsection{Variable-based Interface Formalisms}

In variable-based interface formalisms, such as the formalisms of
\cite{EMSOFT01,CAV02-IM}, communication is mediated by input and
output variables, and the system evolves in synchronous steps.  It is
well known that synchronous, variable-based models can also encode
communication via actions \cite{RM96journal}: the generation of an
output $a!$ is translated into the toggling of the value of an
(output) boolean variable $x_a$, and the reception of an input $a?$ is
encoded by forcing a transition to occur whenever the (input) variable
$x_a$ is toggled.  This encoding is made more attractive by syntactic
sugar \cite{RM96journal}.  However, this encoding prevents the
modeling of many-to-one and many-to-many communication.

In fact, due to the synchronous nature of the formalism, a variable
can be modified at most by one module: if two modules modified it,
there would be no simple way to determine its updated
value.\footnote{A possible way out would be to define that, in case of
simultaneous updates, only one of the updates occurs
nondeterministically.  This choice, however, would lead to a complex
semantics, and to complex analysis algorithms.}  Since the generation
of an output $a!$ is modeled by toggling the value of a boolean
variable $x_a$, this limitation indicates that an output action can be
emitted at most by one module.  As a consequence, we cannot write
modules that can accept inputs from multiple sources: every module
must know precisely which other modules can provide inputs to it, so
that distinct communication actions can be used.  The advance
knowledge of the modules involved in communication hampers module
re-use.

\subsection{Action-based Interface Formalisms}

Action-based interfaces, such as the models of
\cite{FSE01-IA,luca-taormina03,IA-Marktoberdorf04}, enable a natural
encoding of asynchronous communication. However, two interfaces could
be composed only if they did not share output actions.

Furthermore, action-based formalisms lacked a notion of global
variables which are visible to all the modules of a system.  Such
global variables are a very powerful and versatile modeling paradigm,
providing a notion of global, shared state.  Mimicking global
variables in purely action-based models is rather inconvenient: it
requires encapsulating every global variable by a module, whose state
corresponds to the value of the variable.  Read and write accesses to
the variable must then be translated to appropriate sequences of input
and output actions, leading to cumbersome models.

\bibliographystyle{alpha}
\bibliography{main}
\end{document}